\documentclass[article]{aa}
\usepackage{graphicx}
\usepackage[varg]{txfonts}
\usepackage[colorlinks=true,linkcolor=blue, citecolor=blue,filecolor=blue, urlcolor=blue]{hyperref}
\defcitealias{Naujalis2021}{Paper~I}

\begin{document} 

\title{Deriving physical parameters of unresolved star clusters}
\subtitle{VII. Adaptive aperture photometry of the \object{M\,31} PHAT star clusters}
\author{Eimantas Kri{\v s}{\v c}i{\= u}nas \and Karolis Daugevi{\v c}ius \and Rima Stonkut{\.e} \and Vladas Vansevi{\v c}ius}
\institute{Center for Physical Sciences and Technology, Saul{\.e}tekio av. 3, 10257 Vilnius, Lithuania \\
\email{rima.stonkute@ftmc.lt}}
\date{Submitted 9 June 2023; accepted 20 July 2023}

\abstract
{This work is the seventh study in a series dedicated to investigating degeneracies of simultaneous age, mass, extinction, and metallicity determinations of partially resolved or unresolved star clusters with $Hubble$ Space Telescope broadband aperture photometry. In the sixth work (hereafter, Paper~I), it was demonstrated that the adaptive aperture photometry, performed to avoid the majority of the projected foreground and background stars falling within the apertures, gives more consistent colour indices for star clusters.}
{In this study, we aim to supplement the homogeneous multi-colour aperture photometry results published in Paper~I and provide a complete \object{M\,31} Panchromatic $Hubble$ Andromeda Treasury (PHAT) survey star cluster photometry catalogue for further analysis.}
{Following Paper~I, we used a two-aperture approach for photometry. The first aperture is the standard one used to measure total cluster fluxes. The second (smaller) aperture is introduced to avoid the bright foreground and background stars projecting onto the clusters. We selected the radii of smaller apertures to be larger than the half-light radii of the clusters.}
{We present the second part of the star cluster aperture photometry catalogues for a sample of 1477 star clusters from the \object{M\,31} PHAT survey not covered in Paper~I. Compared to the \object{M\,31} PHAT star cluster aperture photometry catalogue published by Johnson et al., adjustments were made to the cluster centre coordinates, aperture sizes, and sky background levels.}
{}

\maketitle

\section{Introduction}
        
The ongoing breakthrough in extragalactic star cluster studies has been made possible by the extensive Panchromatic Andromeda Treasury Program \citep[PHAT;][]{Dalcanton2012} and the Panchromatic Hubble Andromeda Treasury: Triangulum Extended Region \citep[PHATTER;][]{Williams2021} surveys performed with the $Hubble$ Space Telescope (HST). The stellar and cluster populations of \object{M\,31} and \object{M\,33} galaxy disks are aptly represented by published photometry results for a large sample (\citealp[2753 objects:][]{Johnson2015}; \citealp[1214 objects:][]{Johnson2022}) of star clusters that reside in diverse environments, from extremely crowded central parts of the galaxies to their rather sparse outskirts. 

Star clusters are characterised by the following key parameters: age, mass, metallicity, and interstellar extinction. These parameters can be derived using various methods and it is likely that the most accurate method is a direct determination of cluster parameters using colour-magnitude diagrams (CMDs) of their individual stars \citep{Johnson2016, Wainer2022}. However, this method is applicable only to well-resolved and not very old ($\lesssim$300\,Myr in the cases of \object{M\,31} and \object{M\,33}) star clusters since stars below the turn-off point must be measured reliably. Also, the method of integral cluster spectroscopy \citep{Caldwell2009, Caldwell2011, CaldwellRomanowsky2016} is suitable for this purpose and informative; however, only rather massive clusters can be reached in \object{M\,31}. Other promising methods are based on the fitting of observed integrated star cluster magnitudes and colour indices to stochastic theoretical models \citep{Deveikis2008, Fouesneau2010, Fouesneau2014, deMeulenaer2013, deMeulenaer2014, deMeulenaer2015a, deMeulenaer2015b, Krumholz2015, Bialo2019}. These methods look promising for the study of unresolved star clusters to much larger distances than the Local Group galaxies, which could be  mimicked using ground based observations of \object{M\,31} clusters \citep{Kodaira2004, Narbutis2007a, Narbutis2008, Bridzius2008}. However, when stochastic models are used, the accuracy of derived parameters strongly depends on the uncertainties of aperture photometry \citep{Narbutis2007b} and the proper account for projecting background and foreground stars (hereinafter field stars). Moreover, \citet{Beerman2012} showed that excluding bright, evolved cluster members can lead to a more precise derivation of cluster parameters, especially in cases of low-mass star cluster aperture photometry. Therefore, we decided to continue the publication of multi-colour adaptive aperture photometry results to supplement the sample of star clusters presented in the sixth study of the series by \citet[\citetalias{Naujalis2021}]{Naujalis2021} with as many clusters from \citet{Johnson2015} as were available. 

We present star cluster photometry results using the same two-aperture method as in \citetalias{Naujalis2021}: (i) standard aperture approach to measure ‘total’ (T) fluxes and magnitudes and (ii) adaptive aperture approach to measure central parts of star clusters and applying an aperture correction (AC), based on the $F475W$ passband measurements, to other passbands: ‘colour’ (C) fluxes and magnitudes. By selecting C apertures, we were able to avoid the majority of bright field stars and ensure consistent colour indices for star clusters in our sample. We stress that adaptive aperture photometry is reliable only for clusters without strong colour gradients beyond the clusters' half-light radii, which could arise due to gradual variations in cluster stellar populations. 

The structure of the paper is as follows. In Sect.\,\ref{sec:data}, we present a brief description of the data and the star cluster sample. In Sect.\,\ref{sec:photometry}, we briefly discuss the applied photometry procedure. In Sect.\,\ref{sec:results}, we present the results of multi-colour aperture photometry.  In Sect.\,\ref{sec:conclusions}, we present conclusions.

\section{Observations and cluster sample}
\label{sec:data}

\subsection{Observation data}

Our research is based on the HST PHAT survey \citep{Dalcanton2012} data obtained from the $Hubble$ Legacy Archive (HLA)\footnote{\url{https://hla.stsci.edu}}. We used the so-called "Level 2" products that have been processed by the automated HLA pipeline (with bias and dark frames subtracted, flat fielding applied, and all available exposures combined). The star cluster dataset is made of six passbands from three different HST channels: the $F475W$ and $F814W$ passbands are from the Advanced Camera for Surveys (ACS/WFC); while the $F275W$, $F336W$, and $F110W$, as well as $F160W$ passbands are from the Wide Field Camera 3 UVIS (WFC3/UVIS) and Wide Field Camera 3 IR (WFC3/IR) channels, respectively. The number of exposures combined to produce the resulting frames varies among the passbands: the $F475W$ frames are combined from five exposures, the $F814W$ and $F160W$ frames from four exposures, the $F275W$ and $F336W$ frames from two exposures, and the $F110W$ frame is produced from a single exposure.
        
\subsection{Cluster sample}

\begin{figure*}[!h]
\centering
\includegraphics[width=18cm]{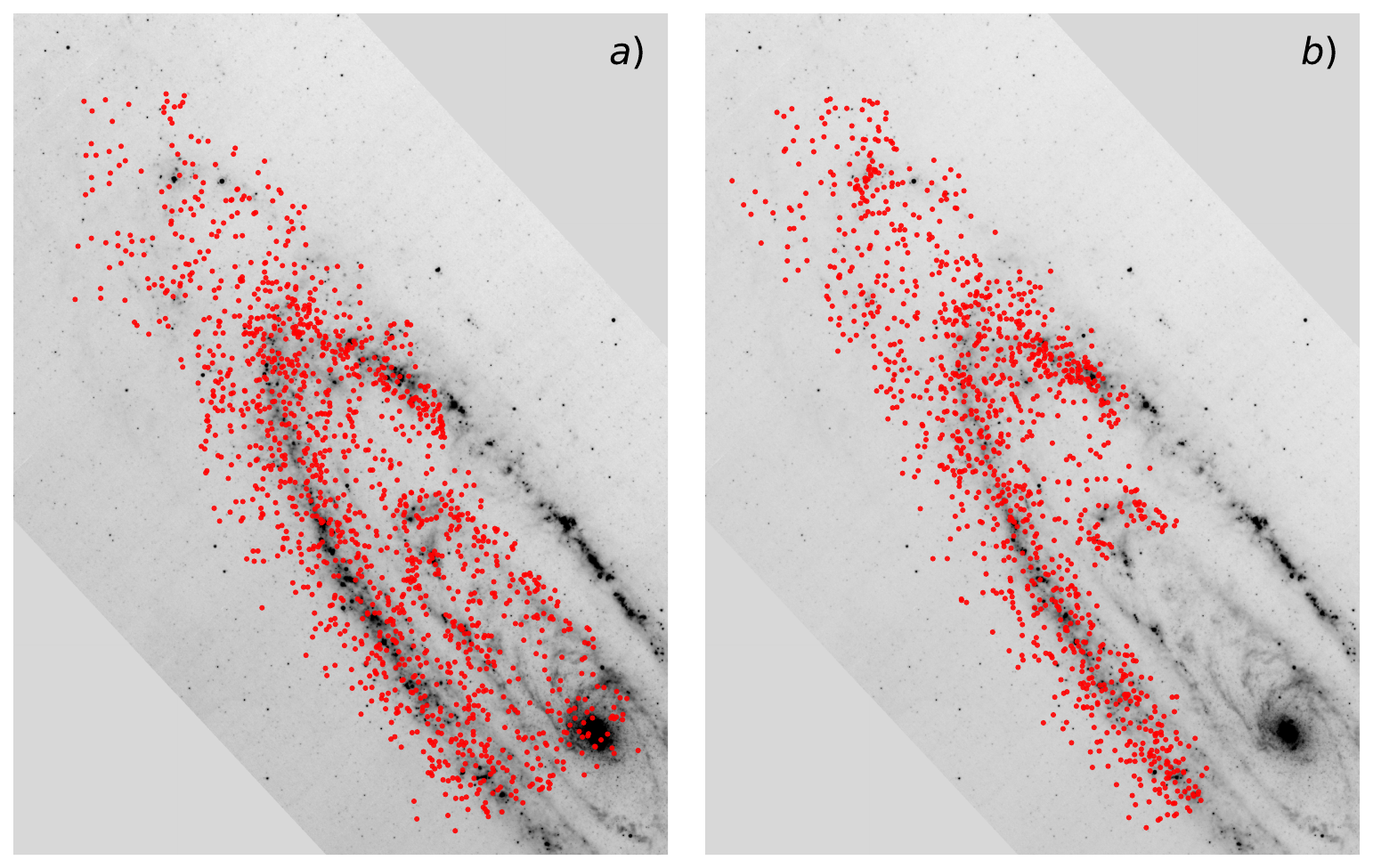}
\caption{Locations of star clusters overlaid on the Multi-Band Imaging Photometer for Spitzer (Spitzer/MIPS) 70~$\mu$m \object{M\,31} map: (a) 1477 clusters measured in this paper; (b) 1181 clusters presented in \citetalias{Naujalis2021}.}
\label{fig:fig1}
\end{figure*}
        
In the \object{M\,31} PHAT fundamental star cluster catalogue published by \citet{Johnson2015}, there are 2753 objects. \citet{Naujalis2021} measured 1181 star clusters by applying the adaptive aperture method; therefore, in this study, we analysed the remaining objects (Fig.\,\ref{fig:fig1}). We visually inspected each object using colour images constructed from the following passbands: $F275W$+$F336W$+$F475W$; $F336W$+$F475W$+$F814W$; $F475W$+$F110W$+$F160W$; and looking into individual grey-scale frames (Fig.\,\ref{fig:fig2}). For photometry, we selected frames with the highest signal-to-noise ratio (S/N) and the lowest number of cosmic ray (CR) artefacts. 

We discarded  90 objects that only have frames in the $F475W$ and the $F814W$ passbands from further analysis. The objects AP1848 and AP2446 were abandoned since they are contaminated by background galaxies. The objects AP1588, AP1661, and AP2687 were also omitted, as they resemble reflections of bright stars or emission nebulae. The frames of the objects AP3630, AP4034, and AP4132 have irreparable empty pixel defects in the $F475W$ passband; therefore, we discarded them. Moreover, three clusters, AP0239, AP1782, and AP3306, seem to be double objects; therefore, we separated them into two parts. In the catalogue (see Sect.\,\ref{sec:results}) the second components are marked: AP60239, AP61782, and AP63306, respectively. We use the names of clusters in the format that they were introduced in, from \citet{Johnson2015}. 

The frames of $F275W$ and $F336W$ passbands contain larger numbers of CR artefacts since they are composed only from the two exposures available, which makes it difficult to reliably clean them in an automated way. To remove CRs, we used the {\tt imedit} task from {\tt IRAF}\footnote{\url{https://iraf-community.github.io}} software \citep{Tody1986}. This enabled us to carefully remove defects projected inside the T apertures. 

There are some clusters residing within the gap between two WFC3/UVIS sensors. Within these gaps, there is a significantly larger number of CRs and higher noise due to only a single exposure being available at that location. Therefore, ultraviolet measurements were discarded where the gap area overlaps with star clusters. Even though the obvious artefacts were removed during visual inspection, in some cases, unrecognized CRs could remain.

In some frames, pixels with zero values (defects) were present. This is an important issue, especially for the $F110W$ passband, in which there is only one exposure available for each field. We assigned an interpolated value for these pixels:\ an average out of eight surrounding non-zero-valued pixels. However, in cases when at least one uncorrected pixel inside the aperture remained after this procedure, we abandoned measurements in this passband.

In the final sample, we measured 1477 star clusters. In Fig.\,\ref{fig:fig1}, locations of the objects analysed in this paper are compared to the distribution of the 1181 clusters from \citetalias{Naujalis2021}. Additionally, using {\tt DS9} software \citep{Joye2003} and based on the $F336W$, $F475W$, and $F814W$ passband frames, we interactively adjusted the centre coordinates of 1331 clusters. In the majority of cases, coordinates were changed only slightly by a few pixels, except for stellar associations, where we selected mostly the cluster-like parts.
        
\section{Aperture photometry}
\label{sec:photometry}
        
\begin{figure*}
\centering
\includegraphics[width=18cm]{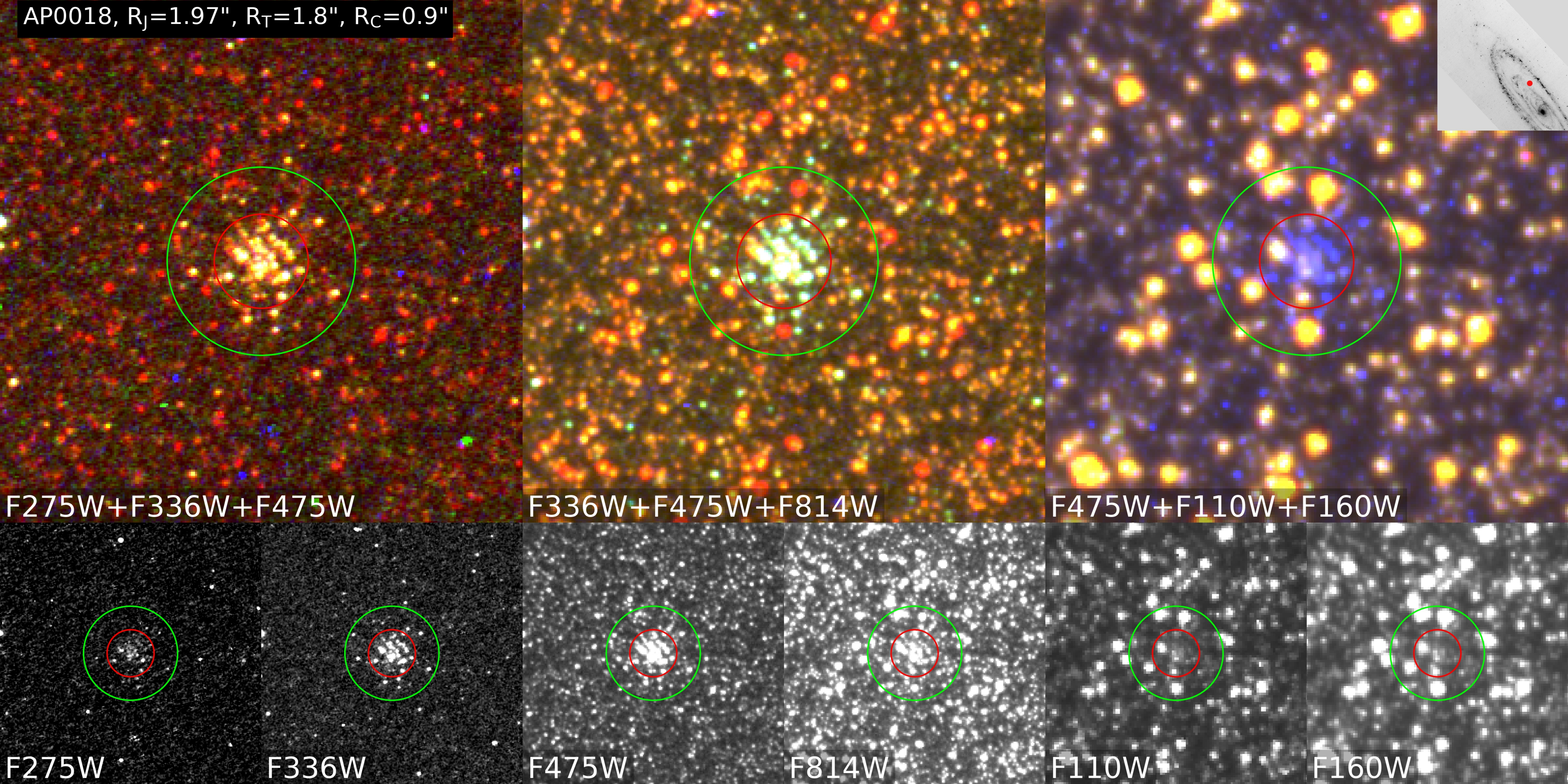}
\caption{Images of the cluster AP0018 in colour panels (top) produced by combining three passbands and grey-scale panels (bottom) produced from individual passband frames (passbands are labelled inside the panels). Green and red circles mark T and C apertures, respectively. The size of each panel is $10\arcsec \times 10\arcsec$. An insert at the top-right corner indicates the location of the cluster in \object{M\,31}. North is up and east is to the left.}
\label{fig:fig2}
\end{figure*}
        
\begin{figure*}[!h]
\centering
\includegraphics[width=18cm]{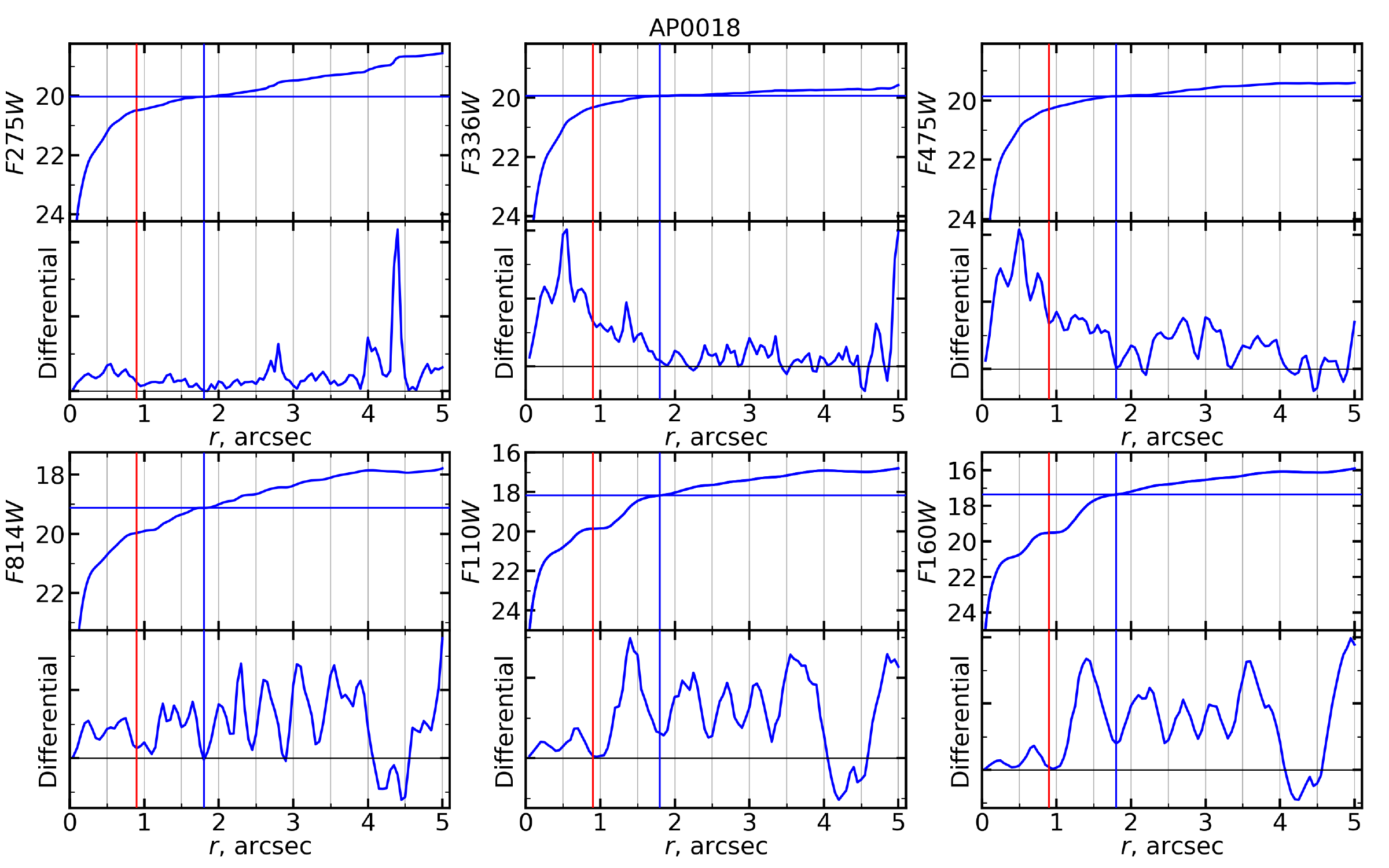}
\caption{Growth curves (top, in magnitudes) and differential flux profiles (bottom, in arbitrary units) for the cluster AP0018. Vertical red and blue lines show C and T aperture radii, respectively. The blue horizontal line indicates the magnitude derived from T fluxes.}
\label{fig:fig3}
\end{figure*}
        
Aperture photometry for each cluster was performed by precisely following the procedures described in \citetalias{Naujalis2021}. A set of figures, shown in Figs.\,\ref{fig:fig2} and\,\ref{fig:fig3} for guidance, were produced for each cluster under consideration and used for the analysis. In Fig.\,\ref{fig:fig2}, the green circle indicates the aperture used to measure the T magnitude of the cluster, while the red circle shows the smaller aperture used to measure the C magnitude, which is more robust for producing consistent colour indices (unbiased by projecting field stars). Figure\,\ref{fig:fig3} shows measured growth curves (top) and differential flux profiles (bottom) in each passband. Vertical red and blue lines mark C and T aperture radii, respectively. Blue horizontal lines indicate magnitudes derived from T fluxes. Negative values in the differential profile indicate that the fluxes in those areas are, on average, lower than the subtracted sky background level. Various positive peaks show the presence of resolved, bright stars and demonstrate the complexity of the surrounding sky backgrounds.

The diverse appearances of star clusters and their surrounding sky backgrounds make it challenging to distinguishing individual stars that are part of the cluster from those that are field objects. However, in the majority of cases, by using the multi-colour star cluster images (Fig.\,\ref{fig:fig2}) together with the cluster profiles (Fig.\,\ref{fig:fig3}), we were able to determine the optimal radii for the T and C apertures.
        
\subsection{Sky background} 
        
Sky backgrounds consist of unresolved stellar and galaxy components, as well as the Milky Way and \object{M\,31} stars projected on the field of view. Therefore, sky backgrounds are unpredictable and their correct determination is the main problem when applying star cluster aperture photometry methods in the case of disk galaxies. Various automated methods have been proposed to determine background levels \citep{BarmbyHuchra2001, KrienkeHodge2007, Johnson2012}. However, their reliability is rather low in extremely dense regions such as the \object{M\,31} disk. Since automatic methods give inconsistent background level results among various passbands, we plotted growth and differential flux profiles in all passbands (Fig.\,\ref{fig:fig3}) for each cluster and derived consistent sky background levels interactively. The main goal of the interactive procedure is to accurately determine the values of unresolved sky background and to estimate the impact of bright resolved field stars. We precisely followed the sky background determination procedure described in detail in \citetalias{Naujalis2021}. 

\subsection{Apertures}
        
To determine T and C magnitudes, we used two co-centred apertures. The main criterion for selecting radii of the T apertures is to avoid splitting into parts the bright star images in the $F336W$, $F475W$, and $F814W$ passbands; the main criterion for selecting radii of the C apertures is to avoid the brightest field stars inside it. It is expected that C magnitudes should provide more consistent cluster colour indices (less contaminated by bright stars), which are critical in determining star cluster parameters from integrated photometry results \citep{deMeulenaer2017}. 

We measured the T magnitudes by applying apertures, the radii (for the majority of objects) of which vary in the range of $\pm$$0\farcs25$ compared with the ones used by \citet{Johnson2015}. In total, we decreased the apertures of 684 clusters and increased the apertures of 733 clusters. The magnitudes of all the star clusters in our sample are measured in at least four passbands (including $F336W$, $F475W$, and $F814W$).

The colour-consistent cluster C magnitudes in all passbands were derived by applying aperture corrections, determined for the $F475W$ passband, ${F475W}_{\rm AC}$. The AC is defined as the difference between magnitudes derived from the fluxes measured through T and C apertures for the $F475W$ passband: ${F475W}_{\rm AC} = {F475W}_{\rm T} - {F475W}_{\rm C}$. We followed the procedure introduced and described in detail in \citetalias{Naujalis2021}.

Here, we want to stress that in some complicated cases the accuracy of the photometry results can be underestimated. However, in most cases, the adaptive aperture photometry provides colour-consistent cluster magnitudes that fit well with the models (see Sect.\,\ref{sec:results}).

\subsection{Photometric uncertainties}
        
\begin{figure*}[!h]
\centering
\includegraphics[width=18cm]{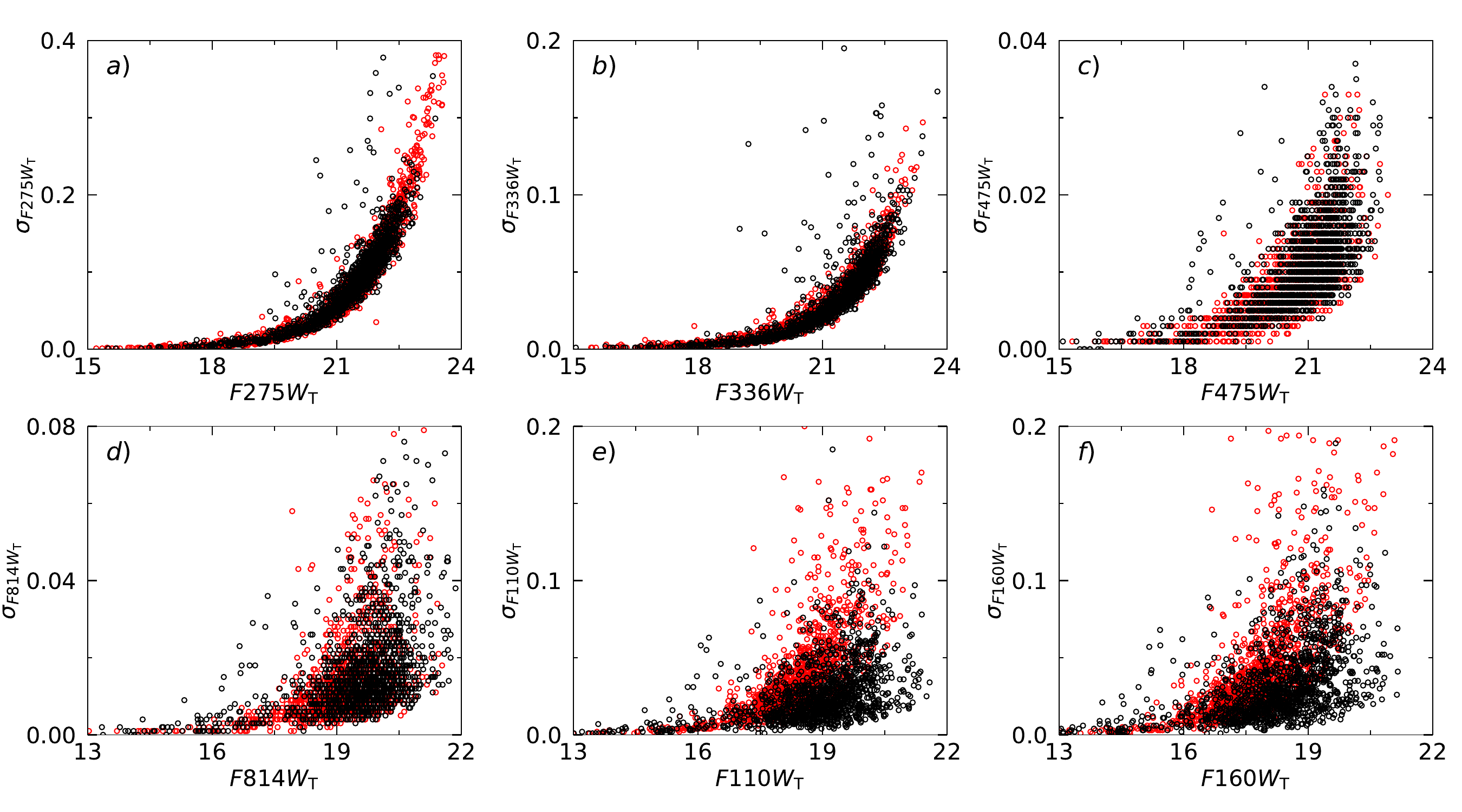}
\caption{Photometric uncertainties of star cluster T magnitudes ($\sigma_{\rm T}$) in all six passbands versus their T magnitudes. The clusters studied in this work are marked by black circles and clusters taken from \citetalias{Naujalis2021} -- by red circles.}
\label{fig:fig4}
\end{figure*}
        
\begin{figure*}[!h]
\centering
\includegraphics[width=18cm]{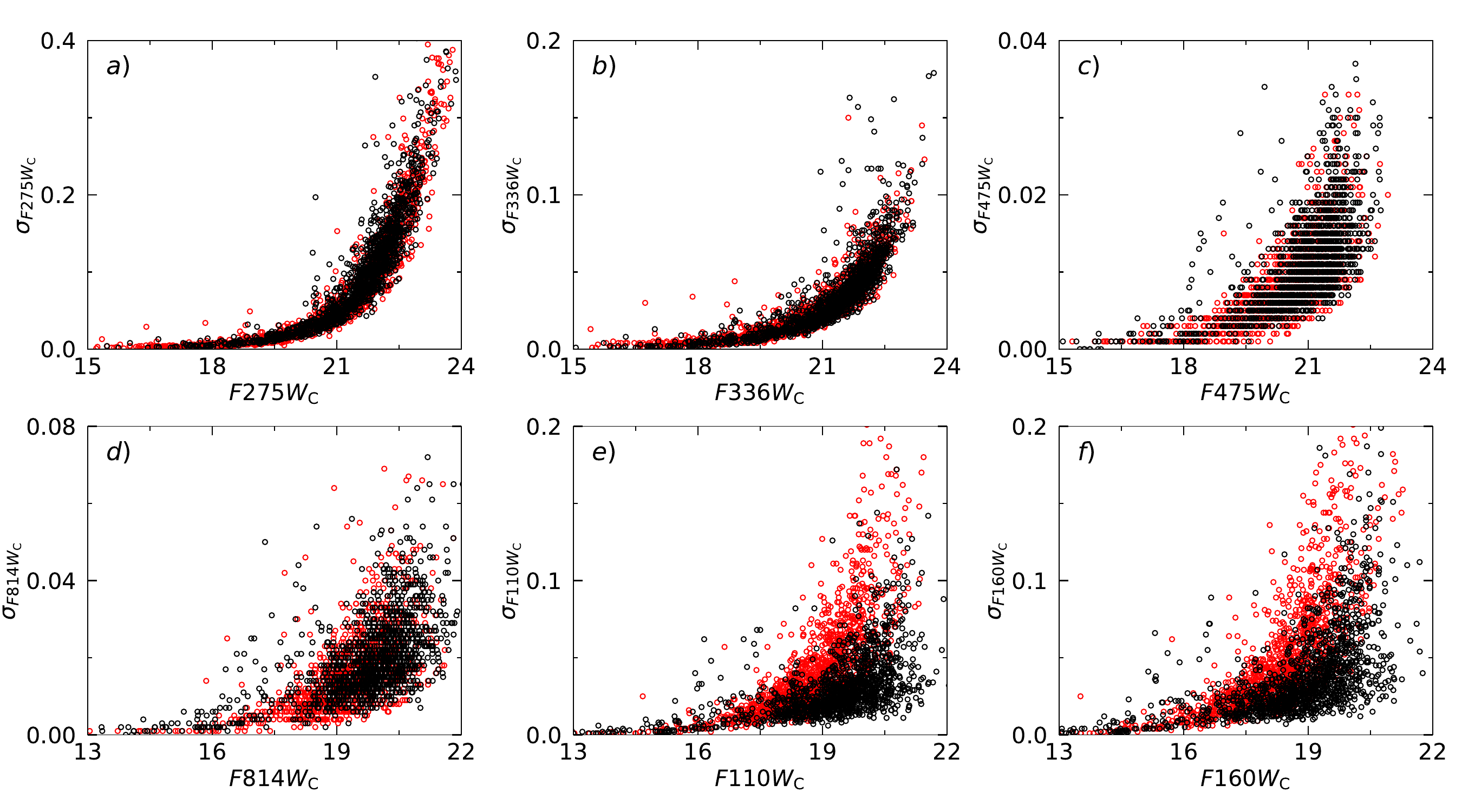}
\caption{Photometric uncertainties of star cluster C magnitudes ($\sigma_{\rm C}$) in all six passbands versus their C magnitudes. The clusters studied in this work are marked by black circles and clusters taken from \citetalias{Naujalis2021} -- by red circles.}
\label{fig:fig5}
\end{figure*}
        
\begin{figure*}[!h]
\centering
\includegraphics[width=18cm]{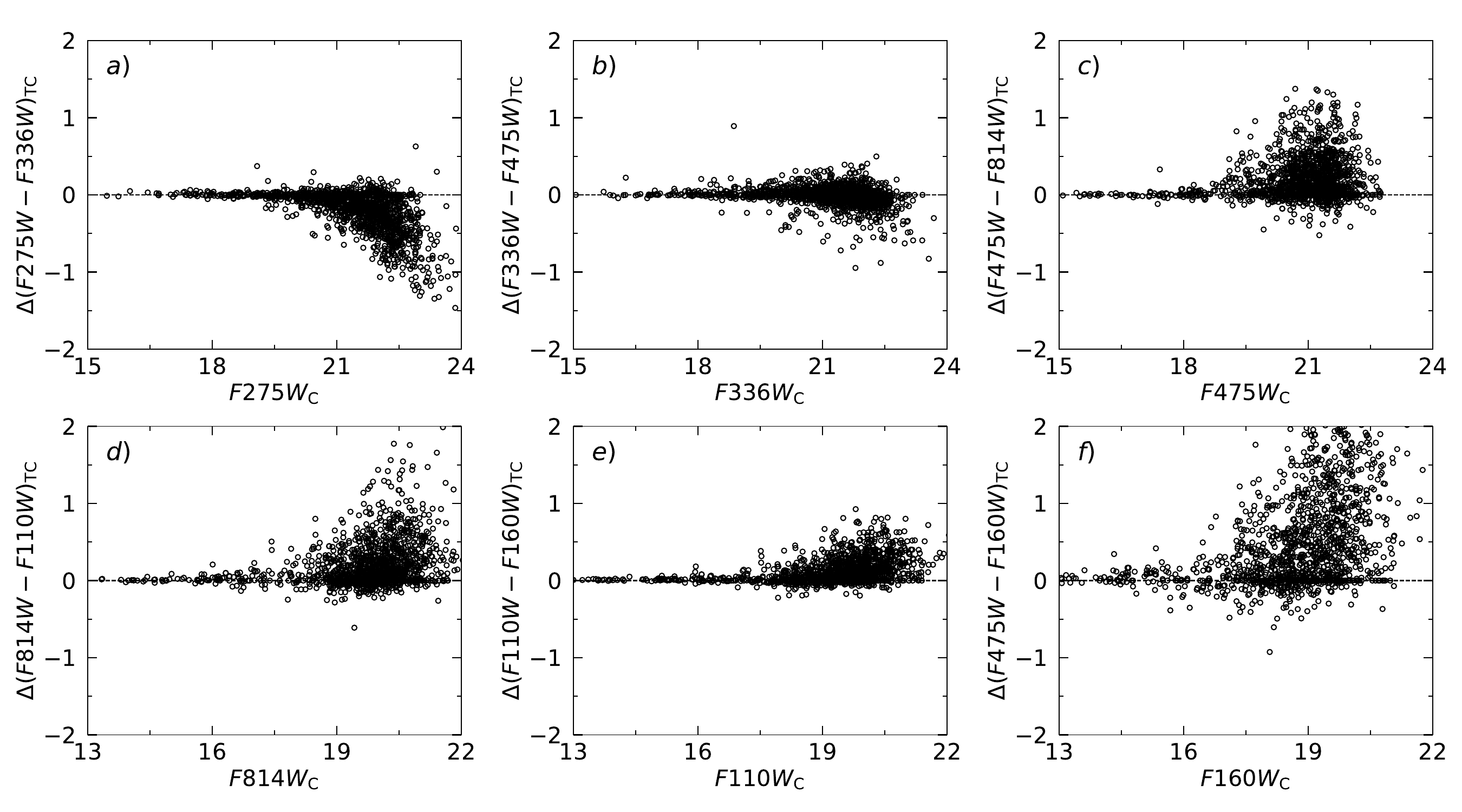}
\caption{Differences between colour indices derived from the T and C magnitudes (the colour index calculated from T magnitudes minus the colour index calculated from C magnitudes).}
\label{fig:fig6}
\end{figure*}
        
In order to estimate photometric uncertainties, we considered two main sources that contribute to the errors of cluster magnitudes. The first contributor of magnitude uncertainty is sky background variation; the second part of errors could arise because of a possible aperture position bias and different sizes of stellar images in various passbands. To solve these problems, we used the same prescription to estimate photometric errors as in \citetalias{Naujalis2021}.

Due to high photometric uncertainties given in \citet{Johnson2015}, star clusters analysed in this study were omitted in \citet{deMeulenaer2017} and \citetalias{Naujalis2021}. However, in this study we demonstrate that adaptive aperture photometry can be applied even to formerly rejected star clusters. Figs.\,\ref{fig:fig4} and\,\ref{fig:fig5} show estimated T and C magnitude uncertainties (black) compared with the uncertainties from \citetalias{Naujalis2021} (red). In both cases, the photometric errors in the UV passbands are dominated by sky background uncertainties and correlate with magnitudes, except for a few dozens of clusters which have a higher scattering due to a low S/N or CR artefact residues. Uncertainties in the IR passbands are spread more widely due to the presence of irregularly distributed bright stars. However, we notice that star clusters in this study reach fainter magnitudes in the IR passbands with photometric errors that are still acceptable,  compared to the data from \citetalias{Naujalis2021}. 
        
\section{Multi-colour photometry results}
\label{sec:results}

\begin{table*}
\caption{\object{M\,31} star cluster T aperture photometry results.}
\label{table:tab1}
\centering
\begin{tabular}{rrrrrrrrrrr}
\hline\hline
AP & R.A.(2000)\tablefootmark{a} & DEC(2000)\tablefootmark{a} & R$_{\rm T}$\tablefootmark{b} & $F275W_{\rm T}$ & $F336W_{\rm T}$ & $F475W_{\rm T}$ & $F814W_{\rm T}$ & $F110W_{\rm T}$ & $F160W_{\rm T}$ \\
\hline
12 & 11.456979 & 41.657342 & 1.95 & 21.201 & 20.491 & 19.887 & 18.384 & 17.784 & 17.124 \\
&&&&0.082 & 0.017 & 0.005 & 0.004 & 0.005 & 0.007\tablefootmark{c} \\
17 & 11.490402 & 42.027507 & 1.80 & 20.857 & 20.724 & 20.377 & 19.523 & 19.326 & 19.066 \\
&&&&0.057 & 0.019 & 0.005 & 0.006 & 0.013 & 0.020 \\
18 & 10.871496 & 41.576741 & 1.80 & 20.035 & 19.946 & 19.860 & 19.128 & 18.171 & 17.364 \\
&&&&0.030 & 0.010 & 0.006 & 0.013 & 0.020 & 0.019 \\
\hline
\end{tabular}
\tablefoot{The table shows an excerpt from data presented in the catalogue. \tablefoottext{a}{The R.A.(2000) and DEC(2000) coordinates are in degrees.} \tablefoottext{b}{The T aperture is in arcseconds.} \tablefoottext{c}{The uncertainties of T magnitudes ($\sigma_{\rm T}$) in corresponding passbands.}}
\end{table*}
        
\begin{table*}
\caption{\object{M\,31} star cluster C aperture photometry results.}  
\label{table:tab2}      
\centering
\begin{tabular}{rrrrrrrrrrr}
\hline\hline
AP & R.A.(2000)\tablefootmark{a} & DEC(2000)\tablefootmark{a} & R$_{\rm C}$\tablefootmark{b} & $F275W_{\rm C}$ & $F336W_{\rm C}$ & $F475W_{\rm C}$ & $F814W_{\rm C}$ & $F110W_{\rm C}$ & $F160W_{\rm C}$ \\ 
\hline
12 & 11.456979 & 41.657342 & 0.80 & 21.533 & 20.539 & 19.887 & 18.547 & 18.083 & 17.543 \\
&&&&0.103 & 0.017 & 0.005 & 0.007 & 0.009 & 0.010\tablefootmark{c} \\
17 & 11.490402 & 42.027507 & 0.95 & 20.898 & 20.660 & 20.377 & 19.715 & 19.595 & 19.406 \\
&&&&0.055 & 0.017 & 0.005 & 0.009 & 0.014 & 0.023 \\
18 & 10.871496 & 41.576741 & 0.90 & 20.055 & 19.896 & 19.86 &  19.534 & 19.423 & 19.081 \\
&&&&0.024 & 0.010 & 0.006 & 0.013 & 0.031 & 0.045 \\
\hline
\end{tabular}
\tablefoot{The table shows an excerpt from data presented in the catalogue. \tablefoottext{a}{The R.A.(2000) and DEC(2000) coordinates are in degrees.} \tablefoottext{b}{The C aperture is in arcseconds.} \tablefoottext{c}{The uncertainties of C magnitudes ($\sigma_{\rm C}$) in corresponding passbands.}}
\end{table*}

\begin{figure*}[!h]
\centering
\includegraphics[width=18cm]{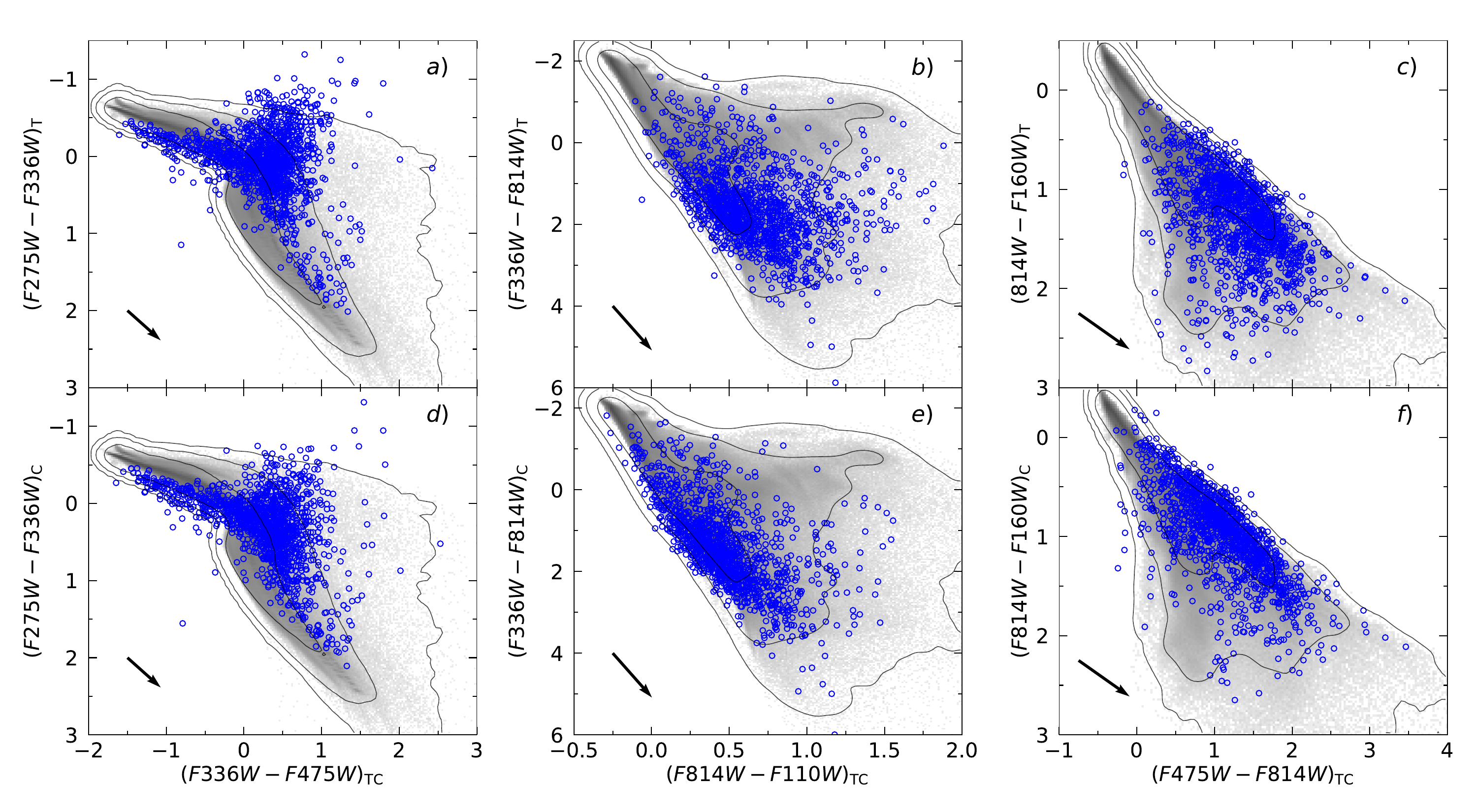}
\caption{Two-colour diagrams showing photometry results of star clusters from this study. Panels a-c show T aperture photometry; panels d-f give the C aperture photometry. The grey colour in the background indicates areas occupied by stochastic star cluster models with masses from $10^2$ to $10^5$\,M$_\odot$. Colour indices indicated on the X axis are constructed from T magnitudes in panels a-c and from C magnitudes in panels d-f. The subscript TC indicates that colour indices are based on T and C aperture photometry in panels a-c and d-f, respectively. Arrows indicate the extinction vectors of $A_V = 1$, assuming the standard Milky Way extinction law.}
\label{fig:fig7}
\end{figure*}

\begin{figure*}
\centering
\includegraphics[width=18cm]{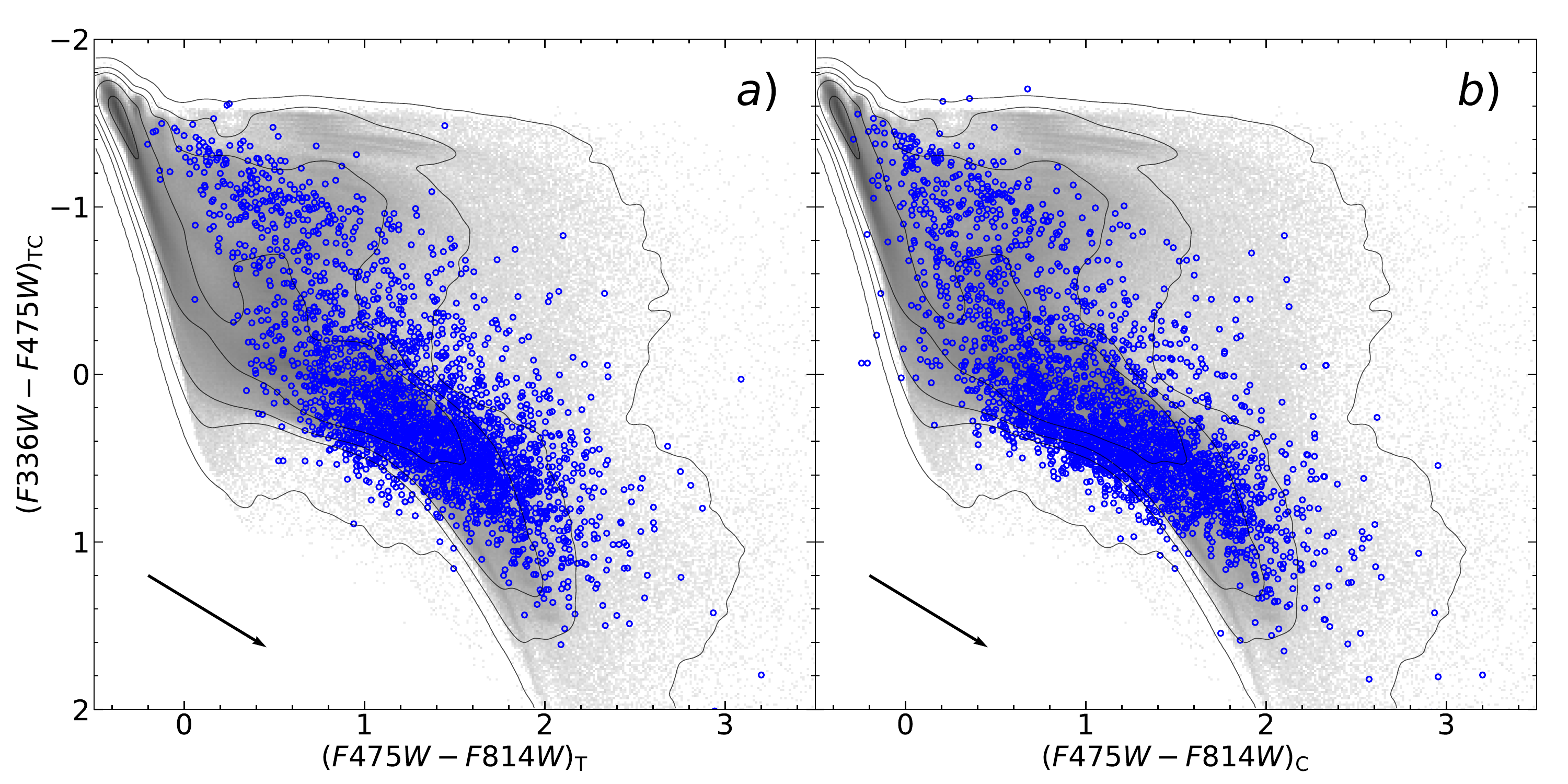}
\caption{Two-colour diagrams showing star clusters from \citetalias{Naujalis2021} and this study over-plotted on the distributions of models. The models of clusters (areas of grey colour) are with masses from $10^2$ to $10^5$\,M$_\odot$. The subscript TC indicates that colour index is based on T and C aperture photometry in panels (a) and (b), respectively. Arrows indicate extinction vectors of $A_V = 1$, assuming the standard Milky Way extinction law.}
\label{fig:fig8}
\end{figure*}
        
The T aperture photometry results, along with the photometric errors ($\sigma_{\rm T}$) in each passband, are provided for all 1477 studied clusters in Table\,\ref{table:tab1}\footnote{Table\,\ref{table:tab1} is available in full at the CDS via anonymous ftp to cdsarc.u-strasbg.fr (130.79.128.5) or via \url{http://cdsarc.u-strasbg.fr/viz-bin/qcat?J/A+A/vol/page}}. The C aperture photometry results with the photometric errors ($\sigma_{\rm C}$) for the same sample of clusters are provided in Table\,\ref{table:tab2}\footnote{Table\,\ref{table:tab2} is available in full at the CDS via anonymous ftp to cdsarc.u-strasbg.fr (130.79.128.5) or via \url{http://cdsarc.u-strasbg.fr/viz-bin/qcat?J/A+A/vol/page}}.

In Fig.\,\ref{fig:fig6}, we show the differences between colour indices derived from T and C magnitudes versus C magnitudes. The differences arise due to the C apertures, which are selected to avoid the brightest field and cluster stars. Other parameters affecting photometry, such as star cluster positions, T aperture sizes, and sky background levels, remain the same in the cases of both apertures. Therefore, the large differences in Fig.\,\ref{fig:fig6} show a high sensitivity to the presence of bright field stars and highlight the importance of a careful consideration of the problems addressed in this study.

In order to test the quality of our photometry data, we compare the results with stochastic star cluster models in the age range of log$_{10}$(t/yr) from 6.6 to 10.1, masses from $10^2$\,M$_\odot$ to $10^5$\,M$_\odot$, and the metallicity, [M/H], range from $-$2.2 to +0.4. These models are based on the PAdova and tRieste Stellar Evolutionary Code (PARSEC)-COLIBRI isochrones\footnote{\url{http://stev.oapd.inaf.it/cgi-bin/cmd}} \citep{Marigo2017} and were computed using the same method as described in \citet{deMeulenaer2017}. Free-of-extinction models are plotted in the background of Figs.\,\ref{fig:fig7} and\,\ref{fig:fig8}. 

In Fig.\,\ref{fig:fig7}, we show two-colour diagrams constructed from T aperture photometry (panels a-c) and C aperture photometry (panels d-f) results compared to the stochastic models (grey in the background). The C aperture photometry results of star clusters follow theoretical models closely for all colour index combinations. In general, the effects of adaptive aperture photometry are equivalent to those discussed in \citetalias{Naujalis2021}. 

The largest differences between the T and C magnitudes are observed in the IR passbands, primarily due to the stronger contamination by numerous red field stars. Photometric results (C magnitudes) of the youngest star clusters are less reliable because measured fluxes are very faint in the IR passbands (Fig.\,\ref{fig:fig7}c,f). The accuracies of UV colour indices (Fig.\,\ref{fig:fig7}a,d) are limited due to the low S/N. Outliers in Fig.\,\ref{fig:fig7}d-f usually have very complicated surrounding sky backgrounds, or have bright field stars falling inside the C aperture. Some star clusters are strongly affected by the presence of interstellar extinction; noticeable systematic shifts in colour indices are evident when comparing observed clusters with the stochastic star cluster models. The extinction vectors, assuming the standard Milky Way extinction law, are shown at the bottom-left corners of panels. 

To show the entire sample of star clusters possessing homogeneous aperture photometry, we merged the catalogues of star clusters in \object{M\,31}, which are presented in this work (1477 objects) and in \citetalias{Naujalis2021} (1181 objects). Figure\,\ref{fig:fig8} shows two-colour diagrams of the entire sample of measured star clusters from the PHAT survey. In Fig.\,\ref{fig:fig8}a, we show our photometry results from applying the T apertures; in Fig.\,\ref{fig:fig8}b:\ photometry results by applying the C apertures. Most clusters align well with the star cluster models drawn in the background (Fig.\,\ref{fig:fig8}). However, the C photometry results, compared with the stochastic star cluster models, are more compatible than the T photometry results.

\section{Conclusions}
\label{sec:conclusions}

We performed multi-colour aperture photometry of 1477 star clusters from the \object{M\,31} galaxy PHAT survey \citep{Dalcanton2012, Johnson2012, Johnson2015}. We used two variants of photometry by applying standard (T) and adaptive (C) apertures. The T and C aperture photometry results for all (1477) clusters are presented in Tables\,\ref{table:tab1} and\,\ref{table:tab2}.

In Figs.\,\ref{fig:fig4} and\,\ref{fig:fig5}, we compared photometry accuracy estimates derived in this work with those from \citetalias{Naujalis2021}. It is evident that uncertainties of the present sample star cluster magnitudes are not higher than in \citetalias{Naujalis2021}. Also, we have demonstrated that the C aperture photometry provides more consistent colour indices compared to the T aperture photometry (Fig.\,\ref{fig:fig7}). 

Finally, in Fig.\,\ref{fig:fig8} we show two-colour diagrams of the presently available entire cluster sample (combined from this study and \citetalias{Naujalis2021}, in total 2658 objects) possessing homogeneous adaptive aperture photometry measurements. Two aspects are notable: our photometry results are in a good fit with the available stochastic cluster models and the T photometry results (Fig.\,\ref{fig:fig8}a) differ significantly compared with the C photometry results (Fig.\,\ref{fig:fig8}b).

We would like to stress two main problems of the star cluster aperture photometry procedure, which limit the accuracy of results in crowded fields: uncertainties on the sky background determination and the projection of field stars into apertures. Therefore, carefully accounting for both problems is the only way to guarantee reliable star cluster aperture photometry results within the disks of such galaxies as \object{M\,31}.

\begin{acknowledgements}
This research made use of the following software: {\tt TOPCAT} (\url{https://www.star.bristol.ac.uk/mbt/topcat}); {\tt SAOImage DS9} (\url{https://ds9.si.edu}); {\tt IRAF} (\url{https://iraf.net}). It is based on observations made with the NASA/ESA $Hubble$ Space Telescope, and obtained from the $Hubble$ Legacy Archive, which is a collaboration between the Space Telescope Science Institute (STScI/NASA), the Space Telescope European Coordinating Facility (STECF/ESA) and the Canadian Astronomy Data Centre (CADC/NRC/CSA). This research was funded by a grant (No. LAT-09/2016) from the Research Council of Lithuania.
\end{acknowledgements}
        
\bibliographystyle{aa}
\bibliography{M31-clusters}

\begin{thebibliography}{32}
\expandafter\ifx\csname natexlab\endcsname\relax\def\natexlab#1{#1}\fi

\bibitem[{{Barmby} \& {Huchra}(2001)}]{BarmbyHuchra2001}
{Barmby}, P. \& {Huchra}, J.~P. 2001,
  \href{http://dx.doi.org/10.1086/323457}{\color{blue}\aj},
  \href{https://ui.adsabs.harvard.edu/abs/2001AJ....122.2458B}{122, 2458}

\bibitem[{{Beerman} {et~al.}(2012){Beerman}, {Johnson}, {Fouesneau},
  {Dalcanton}, {Weisz}, {Seth}, {Williams}, {Bell}, {Bianchi}, {Caldwell},
  {Dolphin}, {Gouliermis}, {Kalirai}, {Larsen}, {Melbourne}, {Rix}, \&
  {Skillman}}]{Beerman2012}
{Beerman}, L.~C., {Johnson}, L.~C., {Fouesneau}, M., {et~al.} 2012,
  \href{http://dx.doi.org/10.1088/0004-637X/760/2/104}{\color{blue}\apj},
  \href{https://ui.adsabs.harvard.edu/abs/2012ApJ...760..104B}{760, 104}

\bibitem[{{Bialopetravi{\v{c}}ius} {et~al.}(2019){Bialopetravi{\v{c}}ius},
  {Narbutis}, \& {Vansevi{\v{c}}ius}}]{Bialo2019}
{Bialopetravi{\v{c}}ius}, J., {Narbutis}, D., \& {Vansevi{\v{c}}ius}, V. 2019,
  \href{http://dx.doi.org/10.1051/0004-6361/201833833}{\color{blue}\aap},
  \href{https://ui.adsabs.harvard.edu/abs/2019A&A...621A.103B}{621, A103}

\bibitem[{{Brid{\v{z}}ius} {et~al.}(2008){Brid{\v{z}}ius}, {Narbutis},
  {Stonkut{\.{e}}}, {Deveikis}, \& {Vansevi{\v{c}}ius}}]{Bridzius2008}
{Brid{\v{z}}ius}, A., {Narbutis}, D., {Stonkut{\.{e}}}, R., {Deveikis}, V., \&
  {Vansevi{\v{c}}ius}, V. 2008,
  \href{http://dx.doi.org/10.48550/arXiv.0902.3167}{\color{blue}Baltic
  Astronomy}, \href{https://ui.adsabs.harvard.edu/abs/2008BaltA..17..337B}{17,
  337}

\bibitem[{{Caldwell} {et~al.}(2009){Caldwell}, {Harding}, {Morrison}, {Rose},
  {Schiavon}, \& {Kriessler}}]{Caldwell2009}
{Caldwell}, N., {Harding}, P., {Morrison}, H., {et~al.} 2009,
  \href{http://dx.doi.org/10.1088/0004-6256/137/1/94}{\color{blue}\aj},
  \href{https://ui.adsabs.harvard.edu/abs/2009AJ....137...94C}{137, 94}

\bibitem[{{Caldwell} \& {Romanowsky}(2016)}]{CaldwellRomanowsky2016}
{Caldwell}, N. \& {Romanowsky}, A.~J. 2016,
  \href{http://dx.doi.org/10.3847/0004-637X/824/1/42}{\color{blue}\apj},
  \href{https://ui.adsabs.harvard.edu/abs/2016ApJ...824...42C}{824, 42}

\bibitem[{{Caldwell} {et~al.}(2011){Caldwell}, {Schiavon}, {Morrison}, {Rose},
  \& {Harding}}]{Caldwell2011}
{Caldwell}, N., {Schiavon}, R., {Morrison}, H., {Rose}, J.~A., \& {Harding}, P.
  2011, \href{http://dx.doi.org/10.1088/0004-6256/141/2/61}{\color{blue}\aj},
  \href{https://ui.adsabs.harvard.edu/abs/2011AJ....141...61C}{141, 61}

\bibitem[{{Dalcanton} {et~al.}(2012){Dalcanton}, {Williams}, {Lang}, {Lauer},
  {Kalirai}, {Seth}, {Dolphin}, {Rosenfield}, {Weisz}, {Bell}, {Bianchi},
  {Boyer}, {Caldwell}, {Dong}, {Dorman}, {Gilbert}, {Girardi}, {Gogarten},
  {Gordon}, {Guhathakurta}, {Hodge}, {Holtzman}, {Johnson}, {Larsen}, {Lewis},
  {Melbourne}, {Olsen}, {Rix}, {Rosema}, {Saha}, {Sarajedini}, {Skillman}, \&
  {Stanek}}]{Dalcanton2012}
{Dalcanton}, J.~J., {Williams}, B.~F., {Lang}, D., {et~al.} 2012,
  \href{http://dx.doi.org/10.1088/0067-0049/200/2/18}{\color{blue}\apjs},
  \href{https://ui.adsabs.harvard.edu/abs/2012ApJS..200...18D}{200, 18}

\bibitem[{{de Meulenaer} {et~al.}(2013){de Meulenaer}, {Narbutis}, {Mineikis},
  \& {Vansevi{\v{c}}ius}}]{deMeulenaer2013}
{de Meulenaer}, P., {Narbutis}, D., {Mineikis}, T., \& {Vansevi{\v{c}}ius}, V.
  2013, \href{http://dx.doi.org/10.1051/0004-6361/201220674}{\color{blue}\aap},
  \href{https://ui.adsabs.harvard.edu/abs/2013A&A...550A..20D}{550, A20}

\bibitem[{{de Meulenaer} {et~al.}(2014){de Meulenaer}, {Narbutis}, {Mineikis},
  \& {Vansevi{\v{c}}ius}}]{deMeulenaer2014}
{de Meulenaer}, P., {Narbutis}, D., {Mineikis}, T., \& {Vansevi{\v{c}}ius}, V.
  2014, \href{http://dx.doi.org/10.1051/0004-6361/201423988}{\color{blue}\aap},
  \href{https://ui.adsabs.harvard.edu/abs/2014A&A...569A...4D}{569, A4}

\bibitem[{{de Meulenaer} {et~al.}(2015{\natexlab{a}}){de Meulenaer},
  {Narbutis}, {Mineikis}, \& {Vansevi{\v{c}}ius}}]{deMeulenaer2015a}
{de Meulenaer}, P., {Narbutis}, D., {Mineikis}, T., \& {Vansevi{\v{c}}ius}, V.
  2015{\natexlab{a}},
  \href{http://dx.doi.org/10.1051/0004-6361/201425121}{\color{blue}\aap},
  \href{https://ui.adsabs.harvard.edu/abs/2015A&A...574A..66D}{574, A66}

\bibitem[{{de Meulenaer} {et~al.}(2015{\natexlab{b}}){de Meulenaer},
  {Narbutis}, {Mineikis}, \& {Vansevi{\v{c}}ius}}]{deMeulenaer2015b}
{de Meulenaer}, P., {Narbutis}, D., {Mineikis}, T., \& {Vansevi{\v{c}}ius}, V.
  2015{\natexlab{b}},
  \href{http://dx.doi.org/10.1051/0004-6361/201526544}{\color{blue}\aap},
  \href{https://ui.adsabs.harvard.edu/abs/2015A&A...581A.111D}{581, A111}

\bibitem[{{de Meulenaer} {et~al.}(2017){de Meulenaer}, {Stonkut{\.{e}}}, \&
  {Vansevi{\v{c}}ius}}]{deMeulenaer2017}
{de Meulenaer}, P., {Stonkut{\.{e}}}, R., \& {Vansevi{\v{c}}ius}, V. 2017,
  \href{http://dx.doi.org/10.1051/0004-6361/201730751}{\color{blue}\aap},
  \href{https://ui.adsabs.harvard.edu/abs/2017A&A...602A.112D}{602, A112}

\bibitem[{{Deveikis} {et~al.}(2008){Deveikis}, {Narbutis}, {Stonkut{\.{e}}},
  {Brid{\v{z}}ius}, \& {Vansevi{\v{c}}ius}}]{Deveikis2008}
{Deveikis}, V., {Narbutis}, D., {Stonkut{\.{e}}}, R., {Brid{\v{z}}ius}, A., \&
  {Vansevi{\v{c}}ius}, V. 2008,
  \href{http://dx.doi.org/10.48550/arXiv.0902.4817}{\color{blue}Baltic
  Astronomy}, \href{https://ui.adsabs.harvard.edu/abs/2008BaltA..17..351D}{17,
  351}

\bibitem[{{Fouesneau} {et~al.}(2014){Fouesneau}, {Johnson}, {Weisz},
  {Dalcanton}, {Bell}, {Bianchi}, {Caldwell}, {Gouliermis}, {Guhathakurta},
  {Kalirai}, {Larsen}, {Rix}, {Seth}, {Skillman}, \&
  {Williams}}]{Fouesneau2014}
{Fouesneau}, M., {Johnson}, L.~C., {Weisz}, D.~R., {et~al.} 2014,
  \href{http://dx.doi.org/10.1088/0004-637X/786/2/117}{\color{blue}\apj},
  \href{https://ui.adsabs.harvard.edu/abs/2014ApJ...786..117F}{786, 117}

\bibitem[{{Fouesneau} \& {Lan{\c{c}}on}(2010)}]{Fouesneau2010}
{Fouesneau}, M. \& {Lan{\c{c}}on}, A. 2010,
  \href{http://dx.doi.org/10.1051/0004-6361/201014084}{\color{blue}\aap},
  \href{https://ui.adsabs.harvard.edu/abs/2010A&A...521A..22F}{521, A22}

\bibitem[{{Johnson} {et~al.}(2016){Johnson}, {Seth}, {Dalcanton}, {Beerman},
  {Fouesneau}, {Lewis}, {Weisz}, {Williams}, {Bell}, {Dolphin}, {Larsen},
  {Sandstrom}, \& {Skillman}}]{Johnson2016}
{Johnson}, L.~C., {Seth}, A.~C., {Dalcanton}, J.~J., {et~al.} 2016,
  \href{http://dx.doi.org/10.3847/0004-637X/827/1/33}{\color{blue}\apj},
  \href{https://ui.adsabs.harvard.edu/abs/2016ApJ...827...33J}{827, 33}

\bibitem[{{Johnson} {et~al.}(2012){Johnson}, {Seth}, {Dalcanton}, {Caldwell},
  {Fouesneau}, {Gouliermis}, {Hodge}, {Larsen}, {Olsen}, {San Roman},
  {Sarajedini}, {Weisz}, {Williams}, {Beerman}, {Bianchi}, {Dolphin},
  {Girardi}, {Guhathakurta}, {Kalirai}, {Lang}, {Monachesi}, {Nanda}, {Rix}, \&
  {Skillman}}]{Johnson2012}
{Johnson}, L.~C., {Seth}, A.~C., {Dalcanton}, J.~J., {et~al.} 2012,
  \href{http://dx.doi.org/10.1088/0004-637X/752/2/95}{\color{blue}\apj},
  \href{https://ui.adsabs.harvard.edu/abs/2012ApJ...752...95J}{752, 95}

\bibitem[{{Johnson} {et~al.}(2015){Johnson}, {Seth}, {Dalcanton}, {Wallace},
  {Simpson}, {Lintott}, {Kapadia}, {Skillman}, {Caldwell}, {Fouesneau},
  {Weisz}, {Williams}, {Beerman}, {Gouliermis}, \& {Sarajedini}}]{Johnson2015}
{Johnson}, L.~C., {Seth}, A.~C., {Dalcanton}, J.~J., {et~al.} 2015,
  \href{http://dx.doi.org/10.1088/0004-637X/802/2/127}{\color{blue}\apj},
  \href{https://ui.adsabs.harvard.edu/abs/2015ApJ...802..127J}{802, 127}

\bibitem[{{Johnson} {et~al.}(2022){Johnson}, {Wainer}, {Torresvillanueva},
  {Seth}, {Williams}, {Durbin}, {Dalcanton}, {Weisz}, {Bell}, {Guhathakurta},
  {Skillman}, {Smercina}, \& {Phatter Collaboration}}]{Johnson2022}
{Johnson}, L.~C., {Wainer}, T.~M., {Torresvillanueva}, E.~E., {et~al.} 2022,
  \href{http://dx.doi.org/10.3847/1538-4357/ac8def}{\color{blue}\apj},
  \href{https://ui.adsabs.harvard.edu/abs/2022ApJ...938...81J}{938, 81}

\bibitem[{{Joye} \& {Mandel}(2003)}]{Joye2003}
{Joye}, W.~A. \& {Mandel}, E. 2003, in Astronomical Society of the Pacific
  Conference Series, Vol. 295, Astronomical Data Analysis Software and Systems
  XII, ed. H.~E. {Payne}, R.~I. {Jedrzejewski}, \& R.~N. {Hook},
  \href{https://ui.adsabs.harvard.edu/abs/2003ASPC..295..489J}{489}

\bibitem[{{Kodaira} {et~al.}(2004){Kodaira}, {Vansevi{\v{c}}ius}, {Bridzius},
  {Komiyama}, {Miyazaki}, {Stonkute}, {{\v{S}}ablevi{\v{c}}iut{\.{e}}}, \&
  {Narbutis}}]{Kodaira2004}
{Kodaira}, K., {Vansevi{\v{c}}ius}, V., {Bridzius}, A., {et~al.} 2004,
  \href{http://dx.doi.org/10.1093/pasj/56.6.1025}{\color{blue}\pasj},
  \href{https://ui.adsabs.harvard.edu/abs/2004PASJ...56.1025K}{56, 1025}

\bibitem[{{Krienke} \& {Hodge}(2007)}]{KrienkeHodge2007}
{Krienke}, O.~K. \& {Hodge}, P.~W. 2007,
  \href{http://dx.doi.org/10.1086/511654}{\color{blue}\pasp},
  \href{https://ui.adsabs.harvard.edu/abs/2007PASP..119....7K}{119, 7}

\bibitem[{{Krumholz} {et~al.}(2015){Krumholz}, {Fumagalli}, {da Silva},
  {Rendahl}, \& {Parra}}]{Krumholz2015}
{Krumholz}, M.~R., {Fumagalli}, M., {da Silva}, R.~L., {Rendahl}, T., \&
  {Parra}, J. 2015,
  \href{http://dx.doi.org/10.1093/mnras/stv1374}{\color{blue}\mnras},
  \href{https://ui.adsabs.harvard.edu/abs/2015MNRAS.452.1447K}{452, 1447}

\bibitem[{{Marigo} {et~al.}(2017){Marigo}, {Girardi}, {Bressan}, {Rosenfield},
  {Aringer}, {Chen}, {Dussin}, {Nanni}, {Pastorelli}, {Rodrigues}, {Trabucchi},
  {Bladh}, {Dalcanton}, {Groenewegen}, {Montalb{\'a}n}, \& {Wood}}]{Marigo2017}
{Marigo}, P., {Girardi}, L., {Bressan}, A., {et~al.} 2017,
  \href{http://dx.doi.org/10.3847/1538-4357/835/1/77}{\color{blue}\apj},
  \href{https://ui.adsabs.harvard.edu/abs/2017ApJ...835...77M}{835, 77}

\bibitem[{{Narbutis} {et~al.}(2007{\natexlab{a}}){Narbutis}, {Brid{\v{z}}ius},
  {Stonkut{\.{e}}}, \& {Vansevi{\v{c}}ius}}]{Narbutis2007a}
{Narbutis}, D., {Brid{\v{z}}ius}, A., {Stonkut{\.{e}}}, R., \&
  {Vansevi{\v{c}}ius}, V. 2007{\natexlab{a}},
  \href{http://dx.doi.org/10.48550/arXiv.0712.3959}{\color{blue}Baltic
  Astronomy}, \href{https://ui.adsabs.harvard.edu/abs/2007BaltA..16..421N}{16,
  421}

\bibitem[{{Narbutis} {et~al.}(2007{\natexlab{b}}){Narbutis},
  {Vansevi{\v{c}}ius}, {Kodaira}, {Brid{\v{z}}ius}, \&
  {Stonkut{\.{e}}}}]{Narbutis2007b}
{Narbutis}, D., {Vansevi{\v{c}}ius}, V., {Kodaira}, K., {Brid{\v{z}}ius}, A.,
  \& {Stonkut{\.{e}}}, R. 2007{\natexlab{b}},
  \href{http://dx.doi.org/10.48550/arXiv.0712.3958}{\color{blue}Baltic
  Astronomy}, \href{https://ui.adsabs.harvard.edu/abs/2007BaltA..16..409N}{16,
  409}

\bibitem[{{Narbutis} {et~al.}(2008){Narbutis}, {Vansevi{\v{c}}ius}, {Kodaira},
  {Brid{\v{z}}ius}, \& {Stonkut{\.{e}}}}]{Narbutis2008}
{Narbutis}, D., {Vansevi{\v{c}}ius}, V., {Kodaira}, K., {Brid{\v{z}}ius}, A.,
  \& {Stonkut{\.{e}}}, R. 2008,
  \href{http://dx.doi.org/10.1086/586736}{\color{blue}\apjs},
  \href{https://ui.adsabs.harvard.edu/abs/2008ApJS..177..174N}{177, 174}

\bibitem[{{Naujalis} {et~al.}(2021){Naujalis}, {Stonkut{\.{e}}}, \&
  {Vansevi{\v{c}}ius}}]{Naujalis2021}
{Naujalis}, R., {Stonkut{\.{e}}}, R., \& {Vansevi{\v{c}}ius}, V. 2021,
  \href{http://dx.doi.org/10.1051/0004-6361/202039306}{\color{blue}\aap},
  \href{https://ui.adsabs.harvard.edu/abs/2021A&A...654A...6N}{654, A6
  (Paper~I)}

\bibitem[{{Tody}(1986)}]{Tody1986}
{Tody}, D. 1986, in Society of Photo-Optical Instrumentation Engineers (SPIE)
  Conference Series, Vol. 627, Instrumentation in astronomy VI, ed. D.~L.
  {Crawford}, \href{https://ui.adsabs.harvard.edu/abs/1986SPIE..627..733T}{733}

\bibitem[{{Wainer} {et~al.}(2022){Wainer}, {Johnson}, {Seth},
  {Torresvillanueva}, {Dalcanton}, {Durbin}, {Dolphin}, {Weisz}, {Williams}, \&
  {Phatter Collaboration}}]{Wainer2022}
{Wainer}, T.~M., {Johnson}, L.~C., {Seth}, A.~C., {et~al.} 2022,
  \href{http://dx.doi.org/10.3847/1538-4357/ac51cf}{\color{blue}\apj},
  \href{https://ui.adsabs.harvard.edu/abs/2022ApJ...928...15W}{928, 15}

\bibitem[{{Williams} {et~al.}(2021){Williams}, {Durbin}, {Dalcanton}, {Lang},
  {Girardi}, {Smercina}, {Dolphin}, {Weisz}, {Choi}, {Bell}, {Rosolowsky},
  {Skillman}, {Koch}, {Lindberg}, {Hagen}, {Gordon}, {Seth}, {Gilbert},
  {Guhathakurta}, {Lauer}, \& {Bianchi}}]{Williams2021}
{Williams}, B.~F., {Durbin}, M.~J., {Dalcanton}, J.~J., {et~al.} 2021,
  \href{http://dx.doi.org/10.3847/1538-4365/abdf4e}{\color{blue}\apjs},
  \href{https://ui.adsabs.harvard.edu/abs/2021ApJS..253...53W}{253, 53}

\end{thebibliography}

\end{document}